\documentclass[osajnl,twocolumn,,showpacs,9pt]{revtex4-1} 
\usepackage{amsmath,amssymb,graphicx}

\newcommand{\faom}{f_{\mathrm{AOM}}}
\newcommand{\faomi}{f_{\mathrm{AOM}}^{(i)}}
\newcommand{\faomf}{f_{\mathrm{AOM}}^{(f)}}
\newcommand{\faomm}{f_{\mathrm{AOM}}^{({\mathrm{min}})}}
\newcommand{\faommax}{f_{\mathrm{AOM}}^{({\mathrm{max}})}}
\newcommand{\fl}{f_{\mathrm{L}}}

\newcommand{\frep}{f_{\mathrm{rep}}}
\newcommand{\fbeat}{f_{\mathrm{beat}}}
\newcommand{\fbeatn}{f_{\mathrm{beat}}^{(0)}}

\begin{document}
\title{A method for independent and continuous tuning of $N$ lasers phase-locked to the same frequency comb}

\author{Will Gunton$^1$}
\author{Mariusz Semczuk$^{1,*}$}
\author{Kirk W. Madison$^1$}
\address{$^1$Department of Physics and Astronomy, University of British Columbia, 6224 Agricultural Road, Vancouver, British Columbia, Canada V6T 1Z1\\
$^{*}$Current address: Institute for Quantum Optics and Quantum Information, Austrian Academy of Sciences, Boltzmanngasse 3, Vienna A-1090, Austria}

\begin{abstract}
We present a method of phase-locking any number of continuous-wave lasers to an optical frequency comb (OFC) that enables independent frequency positioning and control of each laser while still maintaining lock to the OFC. The scheme employs an acousto-optic modulator (AOM) in a double pass configuration added to each laser before its light is compared by optical heterodyne with the comb.  The only requirement is that the tuning bandwidth of the double pass AOM setup be larger than half the OFC repetition rate.  We demonstrate this scheme and achieve an arbitrary frequency tuning precision, a tuning rate of 200~MHz/s and a readout precision at the 1~kHz level.
\end{abstract}

\maketitle

Modern experiments often require multiple lasers, differing in frequencies by tens of THz and referenced to atomic transitions or stable cavities. For example, optical clocks based on $^{88}$Sr$^+$ utilize stabilized coherent sources at 422~nm, 674~nm, 1033~nm and 1092~nm~\cite{Margolis19112004,PhysRevLett.95.033001}. For applications in spectroscopy and related experiments aiming at production of ultracold ground state molecules~\cite{K.-K.Ni10102008,PhysRevLett.113.205301,PhysRevLett.113.255301,PhysRevLett.114.205302,Danzl21022010}, an additional technical difficulty arises from the need to vary the frequency of these stabilized lasers over hundreds to thousands of MHz to characterize \emph{a priori} unknown molecular levels. 
 
Since the development of the self-referenced optical frequency combs (OFC)~\cite{Jones28042000,PhysRevLett.84.5102} it has been possible to frequency stabilize lasers to the resulting ``frequency ruler'' by stabilizing the heterodyne beatnote between the laser and the OFC. Using the broadened spectrum of the OFC provides a wide bandwidth for locking, typically corresponding to the 500 - 1100~nm range for titanium sapphire based systems~\cite{Jones28042000}.  Moreover, non-linear optical processes (including frequency doubling the OFC output when the CW frequency is higher or frequency mixing the OFC output with the CW source itself when the CW frequency is lower) can be used to reliably lock the CW laser even when its frequency has no overlap with the OFC's output spectrum \cite{Mills:08, Mills:09}.  It is worth noting that coherent sources at wavelengths below 500~nm are often based on frequency doubling \cite{Friedenauer_2006} or frequency sum generation~\cite{Berkeland:97}, and therefore their stabilization can be done by referencing the seed lasers before nonlinear mixing.

The frequency $f_{\mathrm{L}}$ of a laser frequency locked to an OFC by stabilizing the beatnote produced in an optical heterodyne with the OFC can be expressed as $f_{\mathrm{L}}=f_{\mathrm{CEO}}+n\times f_{\mathrm{rep}}\pm \fbeat$, where $f_{\mathrm{CEO}}$ is the carrier envelope frequency, $f_{\mathrm{rep}}$ is the repetition rate of the comb and $\pm \fbeat$ refers to the heterodyne beatnote between the laser frequency and the comb tooth to which the laser is locked. 

The most straightforward method to simultaneously achieve absolute frequency stability and wide tunability of a CW laser referenced to an OFC relies on the change of $f_{\mathrm{rep}}$~\cite{Washburn:04,Park:06,Mills:09}, which is multiplied by the mode number $n$ (on the order of $10^6$) of the comb element to which the laser is referenced. This allows for scanning ranges of many GHz, typically limited by the mode-hop-free tuning range of a laser. This method, however, is not suitable for independent referencing and tuning of multiple lasers, as their frequencies are all affected by the change of $f_{\mathrm{rep}}$.

More universal schemes do not alter the reference comb's spectrum. Instead, scanning $f_{\mathrm{L}}$ can be achieved by changing $\fbeat$. This approach, however, is confounded by the degeneracy of the beatnote signals produced by adjacent comb elements as $f_{\mathrm{L}}$ approaches the midpoint between the comb teeth and when $\fbeat$ goes to zero frequency when $f_{\mathrm{L}}$ approaches the frequency of any one comb element. In both situations, special techniques that allow jumping over the ``dead zones'' are required~\cite{Jost:02,Fordell:14,Schibli:05}. Continuous frequency shifting has been demonstrated in~\cite{Benkler:13,Rohde:14} where a part of the comb light is split off and is then sent into an electro-optic modulator used to shift the carrier frequency ($f_{\mathrm{CEO}}$) of the OFC by changing the optical phase of the light during the time between subsequent pulses of the mode-locked laser generating the comb. Possibly the easiest to implement methods belonging to this class rely on the change of the locked laser's frequency using an acousto-optic modulator (AOM) either before or after it is compared by optical heterodyne to the OFC~\cite{endnote}.  An advantage in both cases is that the frequency of the laser can be changed without changing $\fbeat$, and thus the radio-frequency (RF) components of the locking electronics need only work well over a narrow frequency band.  However, the frequency tuning range is limited by the bandwidth of the AOM, such that the scanning range rarely exceeds a few hundred MHz.


\begin{figure}[htbp]
  \centering\includegraphics[width=0.47\textwidth]{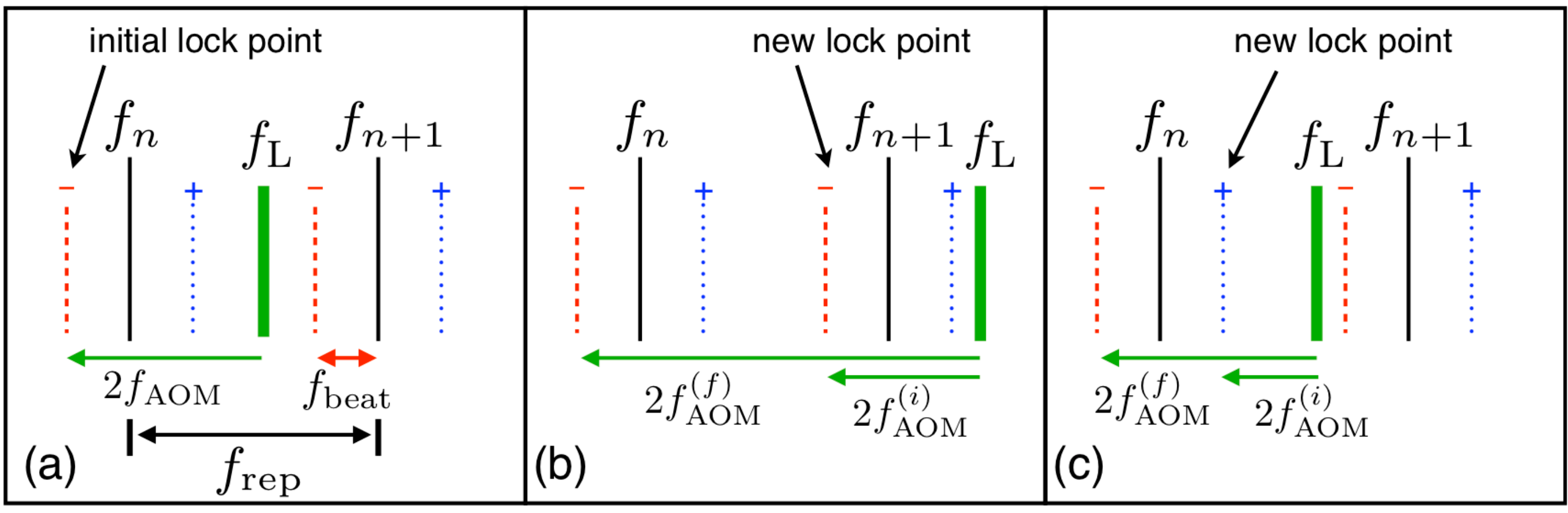}
\caption{(color online) Illustration of the scanning method.  In (a) the laser frequency $\fl$ (short thick green line) is shifted down in frequency by $2\faom$ before heterodyne comparison with the comb element at $f_n$ (thin black line). In (b) the AOM tuning bandwidth is assumed to span $\frep$, and near the high end of the range, the scan is stopped at $\faomf$ and subsequently the AOM frequency is suddenly changed to $\faomi$. Thus $\fbeat$ is unchanged, and the servo re-establishes the lock of $\fl$ referenced to the comb element at $f_{n+1}$.  In (c) the AOM tuning bandwidth is only as large as $\frep/2$, and after scanning to $\faomf$, the scan is stopped and the AOM frequency is changed to $\faomi$ while simultaneously changing the polarity of the error signal so that the servo re-establishes the lock of $\fl$ referenced to the positive beatnote signal (blue dotted line) of the same comb element.}
\label{fig:lock_scheme}
\end{figure}

The locking method we present here, shown in Fig.~\ref{fig:lock_scheme}, achieves continuous and unlimited tuning of the laser output frequency by realizing a ratchet scheme to climb the lock point up or down the comb ladder of frequencies (in analogy with how one might climb a ladder using just one hand). We employ a self-referenced frequency comb generated by an erbium-doped fiber laser, with a center frequency of 
1550~nm~\cite{Mills:08}. The comb light is frequency doubled using a PPLN crystal to match the frequency range of our two tunable cw Ti:sapphire lasers. The details of the optical paths can be found in~\cite{semczuk_thesis} and details of the comb and locking electronics can be found in~\cite{Mills:09}. Our scheme keeps $\fbeat$ constant and employs a double pass AOM to frequency shift the laser light before heterodyne comparison with the OFC.  The locked laser's frequency is therefore $f_{\mathrm{L}}=f_{\mathrm{CEO}}+n\times f_{\mathrm{rep}} + 2 f_{\mathrm{AOM}} \pm \fbeat$ when the AOM is double passed and configured to shift the frequency of the laser down before comparison.  Because the action of the lock electronics is to adjust $f_{\mathrm{L}}$ to keep $\fbeat$ constant, a change of $f_{\mathrm{AOM}}$ results in a change of $f_{\mathrm{L}}$.  The key to our approach that enables the unlimited tuning range is the realization that the speed of frequency changes of the laser output is limited by the finite bandwidth of the servo loop. By exploiting this limitation we can change the comb element to which the laser is locked without disengaging the phase locked loop and without affecting $\fl$. This is accomplished  by a sudden change in $\faom$ from $\faomf$ to $\faomi$ (where 2($\faomf-\faomi) \pm \frep =0$) such that $\fbeat$ remains unchanged and the comb element responsible for producing this beat note is the neighboring element and the integer $n$ has thus changed to $n+1$ or $n-1$.  So long as this change is rapid enough, the servo output will not be disturbed and the servo will simply continue to adjust $\fl$ to keep $\fbeat$ constant through comparison with a different comb element.  Conveniently, the servo will still re-establish the lock to the new comb element even if the sudden frequency change is different from $\frep$, so long as the difference is less than the frequency capture range of the servo.   Then by interlacing sweeps of $\faom$ (slow enough that $\fl$ can follow) with sudden jumps of $\faom$ (to jump the lock point to the next comb element), the output laser frequency can be moved up or down by an arbitrary amount while maintaining the lock to the comb.  The only requirement here is that the tuning bandwidth ($2(\faomm-\faommax)$) of the double pass AOM setup be at least $\frep$.  However, by reversing the polarity of the error signal input to the servo at each jump, one can realize this same scheme with only half this tuning bandwidth - i.e.~with $2(\faomm-\faommax) = \frep/2$.  Here we demonstrate this latter method with an AOM tuning bandwidth of 70~MHz and an OFC with $\frep = 125$~MHz.

Our method is similar in spirit to the solution demonstrated by Biesheuvel et \emph{al.}~\cite{Biesheuvel:13}.  However, we achieve thousand-fold improvement in precision (1~kHz~\cite{PhysRevLett.113.055302} versus 1~MHz) while maintaining a similar frequency scanning rate (200~MHz/s versus 500~MHz/s).  In addition, by virtue of relying on an OFC instead of a Fabry-Perot interferometer (FPI) as a reference, this method has a number of important advantages.  In contrast to a FPI, the action of mode-locking in the OFC automatically guarantees the $\frep$ mode spacing over the entire OFC lasing bandwidth. In addition, locking multiple lasers using an OFC is technically easier than locking them to a FPI.  In the case of the FPI, multiplexing the different sources into a single beam that can be coupled into the FPI requires combining the beams with a cascade of beamsplitters and care must be taken to mode match each beam correctly to only excite a particular transverse spatial mode of the cavity.  Furthermore, the output of the FPI must also be split to distinguish the response for each laser. Locking multiple sources to an OFC is much simpler since the OFC output can be split into multiple reference arms allowing for independent heterodyne comparisons with each of the lasers to be locked without the need for beam or signal multiplexing. While frequency combs are typically used to achieve accuracy and reference cavities are usually employed to provide short term stability, when properly referenced calibrated and stabilized, they can both provide accuracy and stability.
A limitation of our work is that while we achieve with this method phase coherence between the CW lasers~\cite{triplet_data}, we have not demonstrated here that this phase coherence is maintained during the frequency scan at the point when the hand over between modes of the comb is performed. While this hand-over can, in theory, be done in a fully phase coherent way by executing a phase coherent frequency change of the CW laser in a time short compared to the pulse repetition period and timed with the bright interval of the pulse train from the OFC when all of the comb modes have the same optical phase, we did not investigate this possibility in this work.  In applications where phase coherence is required across continuous scans wider than obtained here ($\frep/2$), the EOM based method presented in \cite{Rohde:14} is preferable since the typical update and settling time of an EOM is typically considerably shorter than an AOM.  In all other cases, our method is preferable due to its exceeding simplicity and ease of implementation.  We note that, as suggested in \cite{Rohde:14}, independent and arbitrary phase evolution of the CW optical fields is also possible using this method by controlling the phase of the RF sent to the respective AOMs; however, the bandwidth of this phase evolution is limited by the frequency shifting bandwidth of the AOMs and, as in \cite{Rohde:14}, also by the bandwidth of the phase-locked loop servos.

To demonstrate the feasibility of our method we use two Coherent 899-21 ring cavity Ti:Sapphire lasers (called here TS1 and TS2) with cavity optics suitable for lasing in the 760--830~nm range. A Coherent Verdi V-18 pumps each Ti:Sapphire laser with 9~W of 532~nm light. Laser TS2 is directly frequency offset locked to the comb.  However, the light from the laser TS1 used for the beat note generation is first frequency shifted by an AOM (Intraaction ATD-801A2, with flat diffraction efficiency across a bandwidth of 70~MHz in a double pass configuration) before mixing with the OFC.

Since the tuning bandwidth of our double pass AOM is only slightly larger than $\frep/2$ (62.5~MHz), we realize the ratchet method shown in Fig.~\ref{fig:lock_scheme}c where the polarity of the error signal is inverted for each lock point jump.  The lock error signal is obtained by comparing with a phase/frequency discriminator (based on the Analog Devices AD9901 Chip) the output of a reference oscillator with the RF signal generated by the optical heterodyne of the light from TS1 and the OFC. 

It is important to note that the optical heterodyne of a given CW laser with the comb will produce a series of RF beat frequencies at $\fbeat = \pm \fbeatn + m\frep$ where $m=0,1,2,\ldots$ and $\fbeatn$ is the frequency difference between $\fl$ and the nearest comb element.  Since the OFC light will also mix with itself, the heterodyne detector photocurrent will also contain a series of RF frequencies at $q \frep$ where $q=1,2,3,\ldots$.  To avoid these comb harmonics and equalize the tuning bandwidth required to jump between lock points, one should choose $\fbeat = \pm \frep/4 + m \frep$. We use the third and fifth terms in the beatnote series in order to to reduce the influence of 1/f noise and to ensure that the beatnote frequencies are far from the RF frequencies used to drive the AOMs. The specific choice of the third and fifth terms was due solely to the bandpass filters we had available. Thus, we stabilize TS1 using the heterodyne beatnote at $3\frep - \frep/4 = 343.75$~MHz and TS2 at $5 \frep + \frep/4 =656.25$~MHz.  The integer number $m$ multiplying $\frep$ does not change the ratchet scheme but it does change the comb element responsible for the generation of the signal at $\fbeat$ and thus to which the shifted CW laser is referenced.  This is explained further in the following example.

\begin{figure}[htb]
\centering\includegraphics[width=0.48\textwidth]{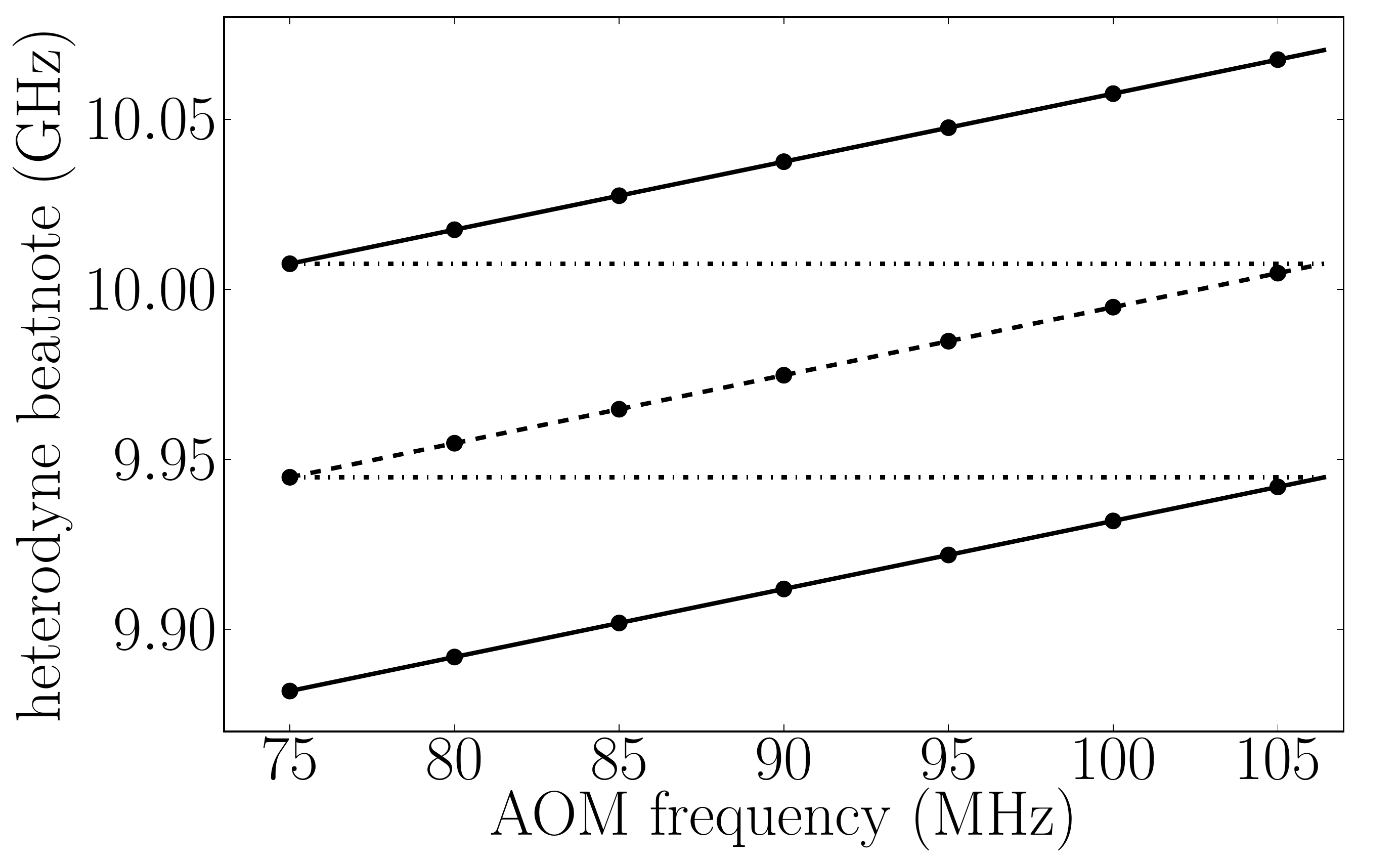}
  \caption{The measured heterodyne beatnote (black dots) between the fixed-frequency laser TS2 and the variable-frequency laser TS1, both phase locked to the OFC, shown as a function of the driving frequency of the double pass AOM. The solid (dashed) lines represent the expected heterodyne beatnote frequency with a negative (positive) servo polarity. The horizontal dashed-dot lines represent the jump of the AOM driving frequency, where the difference in the double passed AOM shift is $\frep/2$ and the resulting $\fbeat$ is unchanged but the comb element responsible for the beatnote changes (see text). For these data, the beatnote was recorded after the 6~ms settling time of the PID servo following a discrete 5 MHz step of the AOM frequency. The optical frequency range shown of this measurement is 190~MHz; however, the full scan range is limited only by the mode-hop-free tuning range of the CW laser.}
\label{fig:beatnote_vs_aom}
\end{figure}

To illustrate the performance of this method, we plot the heterodyne beat note between the fixed frequency laser TS2 and the tunable laser TS1 as a function of the RF frequency driving the AOM (Fig.~\ref{fig:beatnote_vs_aom}). Both lasers were operating at a nominal wavelength of 800~nm, and although the frequency difference shown here is $\approx 10~$GHz (so that the heterodyne beat was visible on a fast photodiode), this method has been used to bridge a 58 GHz gap~\cite{PhysRevLett.113.055302} and an 8 THz gap~\cite{triplet_data}. The limitation on the maximum frequency difference is not due to the locking method itself, but rather the spectrum from the OFC.

Initially, TS1 is locked by stabilizing the heterodyne beatnote at $\fbeat = 3\frep - \frep/4$ with a negative servo polarity.  
By ``negative polarity", we mean that if the laser frequency ($\fl$) increases momentarily, the heterodyne beatnote drops slightly below $\fbeat$ and the servo must respond by decreasing $\fl$. 
The comb element responsible for generating this beat frequency is not the nearest comb element to $\fl$ at $f_n$ but rather the comb element at $f_{n+3}$.
Since TS1 is offset down in frequency by $2\faom$ before comparison, the actual laser frequency is stabilized to $f_{\mathrm{L}}= f_{n+3}+ 2 f_{\mathrm{AOM}} - \fbeat = f_{\mathrm{CEO}}+(n+3) \times \frep + 2 f_{\mathrm{AOM}} - \fbeat = f_{\mathrm{CEO}}+ n \times f_{\mathrm{rep}} + 2 f_{\mathrm{AOM}} + \frep/4$.  Scanning the frequency of TS1 can be done within the bandwidth of the double passed AOM setup either in a continuous manner (by continuously ramping the RF driving frequency of the AOM) or in discrete steps.  For each change of $\faom$, the servo responds by adjusting $\fl$ to keep $\fbeat$ constant.  In our realization, the discrete jumps are limited to be at most $2\Delta f_{\mathrm{AOM}} = 20~$MHz by the bandpass filter bandwidth isolating our RF signal at $\fbeat$.  For such a jump of 20~MHz, the lock is re-established within 6~ms.

In Fig.~\ref{fig:beatnote_vs_aom}, we show the result when $\faom$ is then increased from 75 to 105~MHz in 5~MHz steps.  Since the lock is maintained, $\fl$ increases in 10~MHz steps through a total change of 60~MHz.  As the frequency of TS2 is below that of TS1, the frequency of the heterodyne beatnote between TS1 and TS2 also increases by 60~MHz.  At 105~MHz, the AOM is near its maximum range and it is suddenly changed to 75~MHz while simultaneously changing the polarity of the error signal so that the servo re-establishes the lock of $\fl$ referenced to the positive polarity beatnote signal at $\fbeat$.  In this case, the comb element responsible for this beatnote signal is at $f_{n-2}$.  In this configuration, the laser frequency is stabilized to $f_{\mathrm{L}}=f_{\mathrm{CEO}}+(n-2) \times f_{\mathrm{rep}} + 2 f_{\mathrm{AOM}} + \fbeat = f_{\mathrm{CEO}}+(n+1) \times f_{\mathrm{rep}} + 2 f_{\mathrm{AOM}} - \frep/4 $, and increasing $\faom$ again from 75 to 105~MHz in 5~MHz steps further increases $\fl$ by another 60~MHz.  At the end of this scan, the AOM frequency is suddenly changed from 105 to 75~MHz and the polarity of the error signal is changed again.  At this point, the servo re-establishes the lock of $\fl$ with a negative servo polarity.  Now it is the comb element $f_{n+4}$  that is responsible for the beatnote at $\fbeat$ and the laser is stabilized to $f_{\mathrm{L}}=f_{\mathrm{CEO}}+(n+4) \times \frep + 2 f_{\mathrm{AOM}} - \fbeat = f_{\mathrm{CEO}}+ (n+1) \times f_{\mathrm{rep}} + 2 f_{\mathrm{AOM}} + \frep/4$.  Thus, after two sweeps of $\faom$ interlaced with two rapid jumps by $\frep/4$ and simultaneous servo polarity changes, the laser is now referenced to the $(n+4)^{\mathrm{th}}$ element, one comb element higher than at the beginning of the scan.

By repeating this process, $\fl$ can be moved an arbitrary amount either up or down in frequency while still maintaining the lock to the comb.  As described above, this method exploits the finite response time of the PID controller and allows continuous scanning of the frequency of TS1 over the entire mode-hop-free range while keeping it locked to the comb at all times.  With this system, continuous sweeps including the discontinuous lock point jumps up to 200~MHz/s are possible, and are similar in ramp rate to the 500~MHz/s demonstrated by Biesheuvel et \emph{al.}~\cite{Biesheuvel:13} when locking to a Fabry-Perot cavity. However, for small continuous sweeps using only the double pass AOM (i.e., without the need to switch the polarity of the error signal) sweep rates of $> 3$~GHz/s were obtained. These limitations are set by the inertia of the active elements within the laser (e.g. piezo driven mirrors) and the bandwidth of the locking electronics. However, they are of no (or limited) consequence to experiments focusing on spectroscopy of ultra cold atoms or molecules (where the frequency of the laser must be changed between experimental runs) because the duty cycle of such experiments is typically on the order of seconds.

In conclusion, we have demonstrated a method of stabilizing multiple lasers to a frequency comb without the need for beam or signal multiplexing (as required when using an FPI as a reference) that allows wide, independent, frequency tuning of each laser. The only requirement for each additional CW laser is that a portion of it be frequency shifted in a tunable double passed AOM setup and that it be combined with some portion of light split off from the OFC reference branch. Using an OFC as a reference allows for precise frequency locking of the CW lasers over the entire OFC lasing bandwidth where the $\frep$ mode spacing is guaranteed.  Moreover, non-linear optical processes can be used to reliably lock the CW laser even when its frequency is far from the OFC's output spectrum.  In  addition, the method presented involves maintaining the lock to the comb using a stationary heterodyne beat reference $\fbeat$, and thus the locking electronics are greatly simplified needing only work over a very narrow frequency band.  This approach can be readily used for Raman spectroscopy of widely separated atomic or molecular levels and, if the linewidth of the lasers locked to the comb are narrowed to $\sim$1~kHz level, for the production of ground state polar molecules using stimulated Raman adiabatic passage (STIRAP).

The authors acknowledge financial support from the Natural Sciences and Engineering Research Council of Canada (NSERC / CRSNG), and the Canadian Foundation for Innovation (CFI).  This work was done under the auspices of the Center for Research on Ultra-Cold Systems (CRUCS).  We thank Takamasa Momose and Mark G. Raizen for the long term loan of the Ti:Sapphire lasers and W.~Bowden and G.~Polovy for experimental assistance.

%
%

\end{document}